# Deep2Lead: A distributed deep learning application for small molecule lead optimization

Tarun Kumar Chawdhury *, David J. Grant , and Hyun Yong Jin*

College of Computing, Georgia Institute of Technology Atlanta, GA, USA

**Abstract** Lead optimization is a key step in drug discovery to produce potent and selective compounds. Historically, *in silico* screening and structure-based small molecule designing facilitated the processes. Although the recent application of deep learning to drug discovery piloted the possibility of their *in silico* application lead optimization steps, the real-world application is lacking due to the tool availability. Here, we developed a single user interface application, called Deep2Lead. Our web-based application integrates VAE and DeepPurpose DTI and allows a user to quickly perform a lead optimization task with no prior programming experience.

**Summary:** Deep2Lead allows a user to execute a drug lead optimization pipeline from a single user interface with molecule generation using a Variational Autoencoder (VAE) model and drug target interaction prediction using Deep Purpose library.

**Availability and implementation:** https://deep2lead.dlyog.com/

**Developer and Maintainer Contact:** tarunchawdhury@gatech.edu

## Introduction

Lead optimization is a key step of small molecule drug discovery to produce more potent and selective compounds with pharmacologically acceptable properties. Typically, this process requires a series of investigations of structure-activity relationship (SAR) around each core compound structure, rational optimization of the compounds and iterative assay result interpretation. However, this conventional process is time and resource consuming and does not guarantee to achieve target potency, which is often low-nM level. To overcome this issue, computational methods for *de novo* small molecule design and structure-based, virtual screening including *in silico* molecular docking strategy has been applied (Rester, 2008). However, such an *in silico* application of lead optimization based on deep learning algorithm has been significantly lagged due to the lack of the tool availability.

DeepPurpose is a software library created by Sun group (Fu et al., 2020) and represents a recent advancement in the use of Deep Learning for Drug Target Interaction (DTI). DeepPurpose can serve as a key component of a lead optimization pipeline. Here, we developed a single user interface application, Deep2Lead, which integrates an additional molecule generation step and the power of DeepPurpose DTI. Our application allows a user to quickly perform a lead optimization task with no programming experience required.

**Implementation**

There are three main components: Graphical User Interface (GUI), Variational Autoencoder (VAE) and Drug Target Interaction Predictor (DTI) (Aspuru-Guzik, 2017). The GUI is a web application implemented in HTML / JavaScript and Java EE technologies and allows the user to submit requests to an application programming interface (API). On the server side, the request is passed through the VAE – DTI pipeline and returns a list of valid molecules generated and their predicted binding affinity to the user-specified target (amino acid sequence). Figure 1 shows high level solution design of the application.

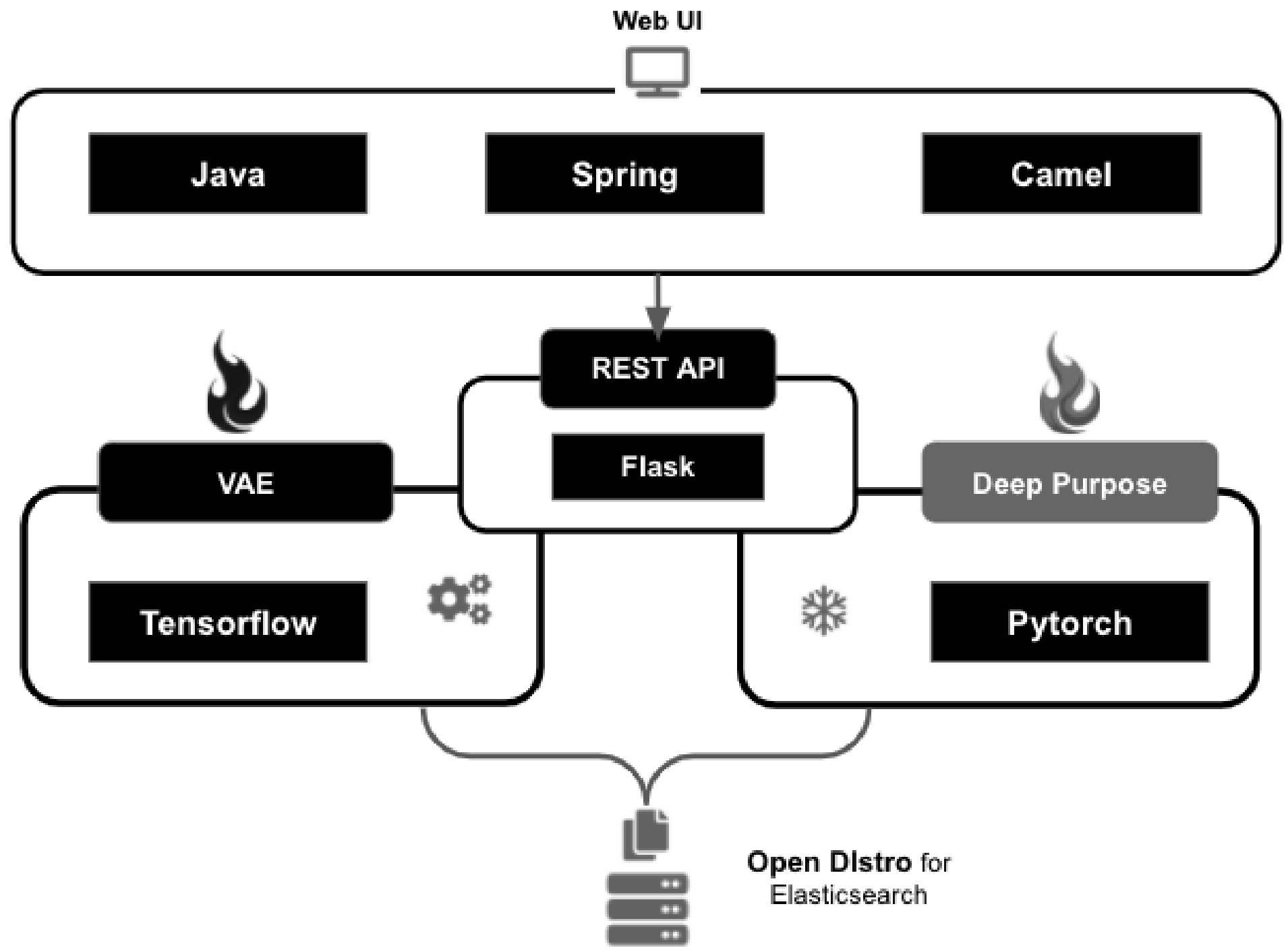

**Figure 1**. High level solution design for Deep2Lead distributed application.

The purpose of the VAE is to generate candidate molecules which are slight variations of the input molecule (SMILE string representation), and is implemented using a pre-trained TensorFlow (Gómez-Bombarelli et al., 2018). These candidate molecules are passed to the DTI module which predicts binding affinity to the user specified target using DeepPurpose API (Fu et al, 2020). The results are returned to the GUI in a sortable table. Both the VAE and DTI (using DeepPurpose) are implemented as provided from their reference papers, and hence their individual efficacy is not assessed here. However, example molecule-target scenarios and results from the Deep2Lead combined pipeline are shown in the supplementary information.

## Usage

Deep2Lead allows a user to generate molecules and assess binding affinity against a particular target. The process is as follows:

1. Select a Drug Target (enter applicable Amino Acid sequence)
2. Select a Lead Compound (specified as SMILE string)
    - Option 1. Paste a SMILES string directly.
    - Option 2. Deep2Lead interface provides a Molecular Designer that users can directly draw chemical structures and convert them into SMILES strings.
3. Enter additional trial parameters (number of attempts, sampling noise parameter)
4. View generated molecules based sorted increasing by predicted IC50

The user can examine the predicted IC50 binding affinity and compare against the lead compound. A specific use case example using a SMILES string of FDA-approved anti-cancer drug Imatinib (Gleevec) and full amino acid sequences of it's known therapeutic target, ABL-BCL fusion protein is shown in **Figure 2**.

**(a)**

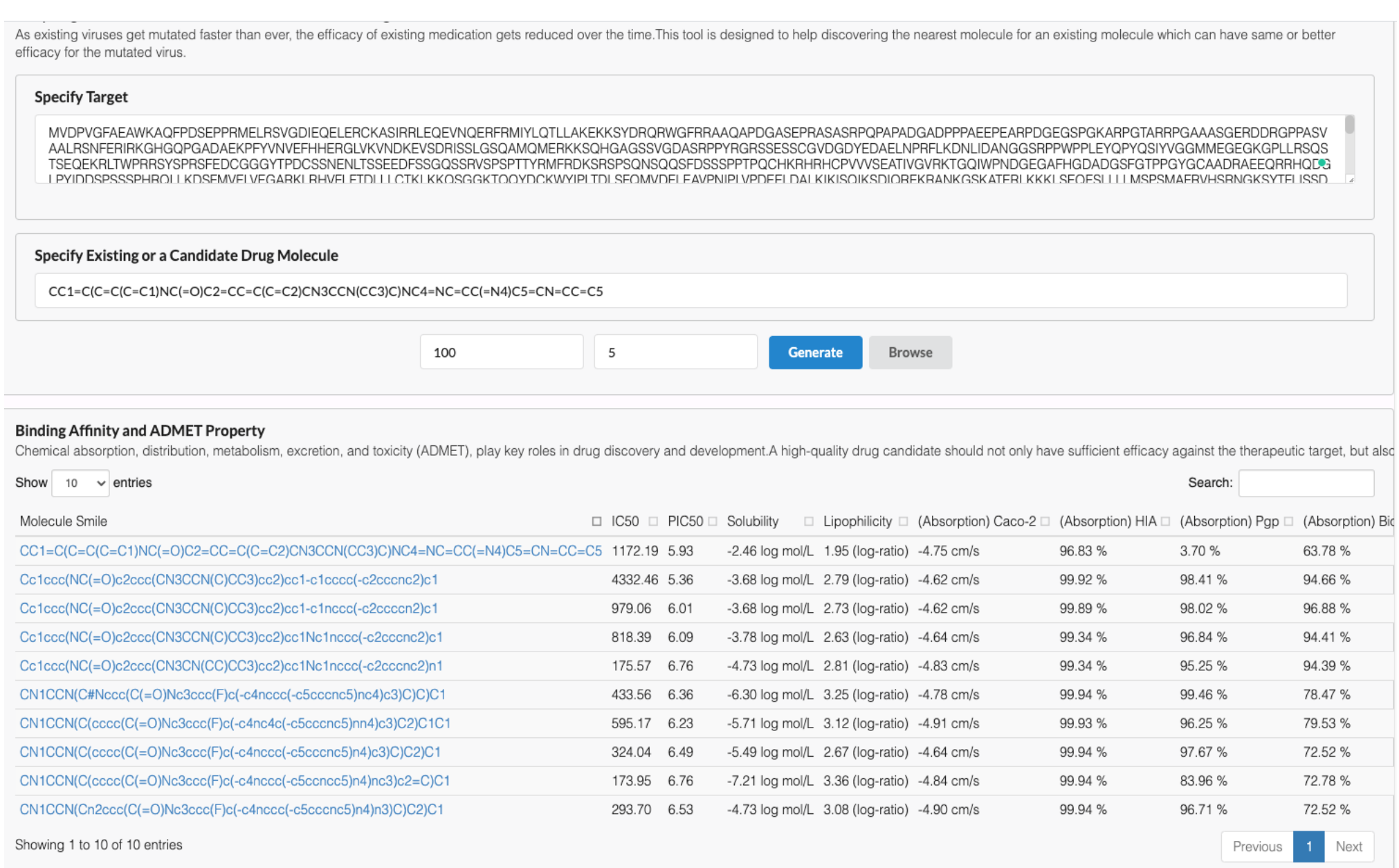

**(b)**

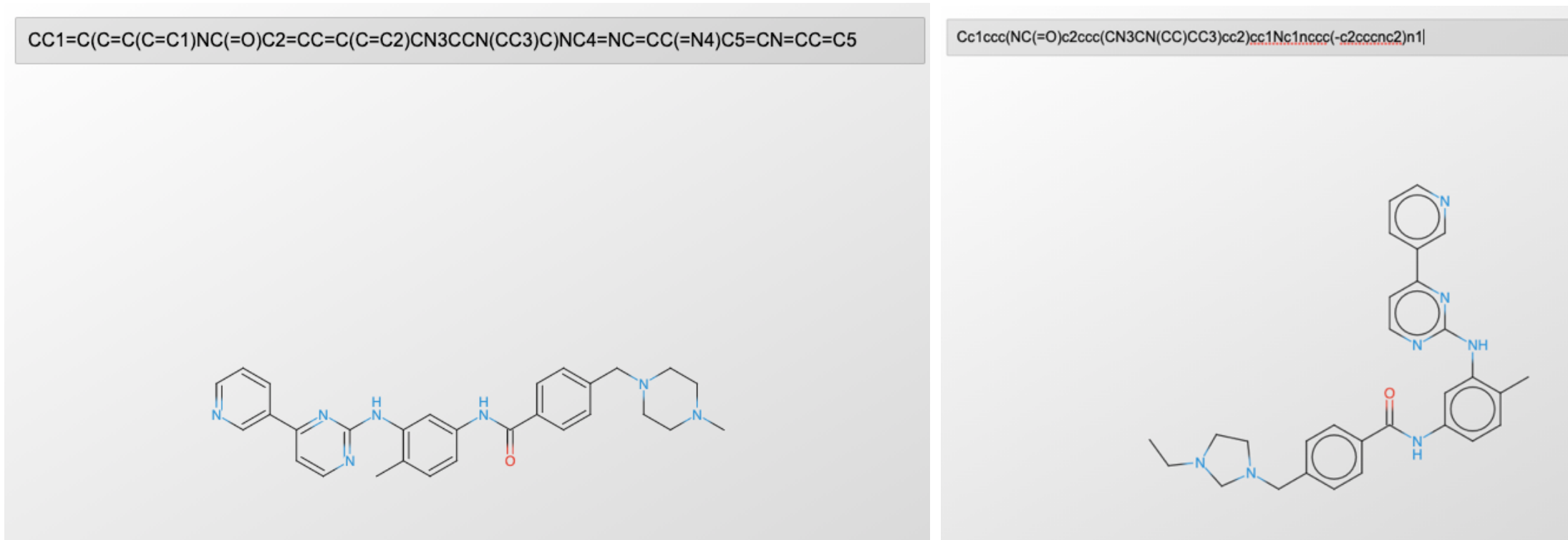

Figure 2. (a) Use case of FDA approved anti-cancer agent Imatinib (Gleevec) and its known target, BCR-ABL fusion protein. Using Imatinib as a lead molecule, we attempt to generate novel candidate inhibitors that potentially have better pharmacological properties than Imatinib. As expected, Deep2Lead was able to generate a small molecule that is 10 times more potent than Imatinib (using Deep Purpose Parameters: Affinity Prediction Model Type = CNN-CNN, ADMET Prediction model Type = CNN). (b) by clicking each molecule, users can easily explore the structural information of the newly generated molecules, including original input (left) and the molecule of interest (right).

## Conclusions

Deep2Lead successfully completes the task of integrating VAE molecule generation and DTI prediction into a single user interface. This allows users with limited programming experience to leverage complex deep-learning models and complete a full lead optimization pipeline. The resulting molecule list can be used for further practical validation of drug-target interaction. While Deep2Lead can generate molecules and predict binding affinity to a target, there is no guarantee that generated molecules will have improved properties compared to the original input. This is due to the nature of the random sampling process used in the VAE. Future work could explore optimization techniques to efficiently search the latent space for candidate molecules based on the DeepPurpose predicted binding affinity.